\documentclass[epj,final]{svjour}
\usepackage{latexsym}
\usepackage[dvips]{epsfig}
\usepackage[dvips]{graphicx}

\newcommand{\sid}{\!} 
\newcommand{\sdf}{\,} 

\newcommand{\wideop}[1]{\,#1\,} 

\newcommand{\rd}{\mathrm{d}} 

\newcommand{\order}[1]{\mathcal{O}\round{#1}}

\newcommand{\erfi}{\mathrm{erfi}}

\newcommand{\sign}{\mathrm{sign}}

\renewcommand{\Im}{\mathrm{Im}}

\newcommand{\shb}[2]{\hspace{#1}\mbox{#2}\hspace{#1}} 

\newcommand{\lrarrowsubtext}[1]{\stackrel{\longrightarrow}{\mbox{\tiny $#1$}}}

\usepackage{amsbsy}
\usepackage{amssymb}

\newcommand{\GF}[2]{\Psi_{#1}^{#2}} 
\newcommand{\GFa}[2]{\bar{\Psi}_{#1}^{#2}} 
\newcommand{\GFaGF}[2]{\bar{\Psi}_{#1}^{#2}\Psi_{#1}^{#2}} 

\newcommand{\GS}[2]{\mathcal{S}_{#1}^{#2}} 

\newcommand{\pathintPsi}{\int \sid \mathcal{D}\,\Psi}

\newcommand{\pathintQ}{\int \sid \mathcal{D}\,Q}

\newcommand{\yzfull}[1]{y_{#1 0}} 
\newcommand{\ymfull}[3]{y_{#2 #3}^{#1}}
\newcommand{\Hzfull}[1]{H^0_{#1}}
\newcommand{\Hmfull}[2]{{H^{#2}_{#1}}}

\newcommand{\ym}[2]{y^{#1}_{#2}}
\newcommand{\Hz}{H_0}

\newcommand{\V}[2]{{V^{#2}_{#1}}}

\newcommand{\sumii}[1]{\sum_{#1 =-\infty}^\infty} 

\newcommand{\intcond}[1]{\int_{#1}} 

\newcommand{\multsum}[2]{\sum_{#1 \atop #2}}
\newcommand{\multprod}[2]{\prod_{#1\atop #2}}

\newcommand{\intGauss}[1]{\int_{#1}^G}

\newcommand{\tr}[1]{\mathrm{Tr}_{#1}}

\newcommand{\round}[1]{\left( #1 \right)}

\newcommand{\squared}[1]{\left[ #1 \right]}

\newcommand{\absolute}[1]{\left| #1 \right|}

\newcommand{\expbb}[1]{\exp\left(#1\right)} 

\newcommand{\avth}[2]{{\left\langle #2 \right\rangle}^{#1}} 
\newcommand{\avdo}[2]{{\left[ #2 \right]}_{J}^{#1}} 
\newcommand{\avAloc}[2]{{\left\langle #2 \right\rangle}_{{\mathcal{A}_{\mbox{\tiny eff}}^{\mbox{\tiny CPA}}}}^{#1}} 

\newcommand{\avyint}[1]{\left[#1\right]_{\mathbf{y}}^{(z)}}
\newcommand{\avyintzo}[1]{\left[#1\right]_{\mathbf{y}}^{(0)}}
\newcommand{\avystaticint}[1]{\left[#1\right]_{\ym{}{0}}^{(z)}}
\newcommand{\avzint}[1]{\intGauss{z}\sdf#1}

\newcommand{\weightP}{W}

\newcommand{\AzeroT}{\bar{A}}

\newcommand{\ts}{\tilde{t}} 
\newcommand{\Js}{\tilde{J}} 
\newcommand{\ssp}{{\sigma\sigma'}}

\newcommand{\se}[1]{\Sigma_{#1}} 

\newcommand{\replim}{\lim_{n\to 0}} 

\newcommand{\lc}{l_c}

\newcommand{\Qabfull}{Q^{\tau\tau'}_{\alpha\beta}}
\newcommand{\Qaa}{Q^{\tau\tau'}_{\alpha\alpha}}
\newcommand{\Qab}{Q^{\tau\tau'}_{\alpha\neq\beta}}

\newcommand{\qt}{{\tilde{q}}}
\newcommand{\ttp}{{\tau\tau '}}
\newcommand{\dt}{{(\tau-\tau')}}

\newcommand{\Az}{\mathcal{A}_0} 
\newcommand{\AJ}{\mathcal{A}_J} 
\newcommand{\At}{\mathcal{A}_t} 
\newcommand{\Aeff}{\mathcal{A}_{\mbox{\footnotesize eff}}} 

\newcommand{\aneqb}{\alpha\neq\beta}
\newcommand{\ia}{i\alpha}

\newcommand{\Tfree}{T_{0}} 

\newcommand{\Gloc}[2]{\Gamma^{#2}_{#1}}

\newcommand{\Gmat}[1]{\tens{G}_{#1}}
\newcommand{\Glocmat}[1]{\tens{\Gamma}_{#1}}
\newcommand{\Tmat}{\tens{T}}
\newcommand{\Vmat}[1]{\tens{V}_{\!\!#1}}

\newcommand{\refeqn}[1]{(\ref{#1})}
\newcommand{\refsec}[1]{\ref{#1}}

\newcommand{\e}{\mathrm{e}}
\newcommand{\smsp}{\,}
\newcommand{\seco}{self-consisten\-cy }

\begin{document}
\title{Dynamical CPA approach to an itinerant fermionic spin glass model}
\author{M. Bechmann \and R. Oppermann 
}                     
%
%
\institute{Institut f\"ur Theoretische Physik und Astrophysik, Universit\"at W\"urzburg, D-97074 W\"urzburg, Federal Republic of Germany }
\date{Received: date / Revised version: date}
%
\abstract{
We study a fermionic version of the Sherrington-Kirkpatrick model including nearest-neighbor hopping on a $\infty$-dimensional simple cubic lattices.
 The problem is reduced to one of free fermions moving in a dynamical effective random medium. By means of a CPA method we derive a set of self-consistency equations for the spin glass order parameter and for the Fourier components of the local spin susceptibility. 
In order to solve these equations numerically we employ an approximation scheme which restricts the dynamics to a feasible number of the leading Fourier components. From a sequence of systematically improved dynamical approximations we estimate the location of the quantum critical point.

} 
\titlerunning{Dynamical CPA approach to an itinerant fermionic spin glass model}
\maketitle
\section{Introduction}
\label{intro}
Quantum-dynamical mean field theories are highly appreciated, helpful methods to
provide insight into models of strongly interacting systems \cite{geor_etal_96_DMFT}.
A well-known case is certainly the Hubbard model including its various model ramifications.
Fermionic spin glass models, though describing the quite different physics of randomly
interacting disordered systems, belong to this category as well; in particular,
they also deal with an intense interplay between magnetism and transport properties.

Quantum dynamics in systems with random many-body interactions are known to contain a
challenging technical difficulty which arises from the double-time dependence of the
spin glass field.
This feature of quantum spin dynamics is hard to handle even in the
dynamical mean field theory (DMFT) and requires further simplifications, such as the limit $M\to\infty$ in $SU(M)$-generalizations of spin systems \cite{parc_geor_99}.

In earlier work the so-called static approximation (defined by a static effective
spin glass field) was used to construct a systematic low temperature expansion
for metallic spin glasses \cite{op_bi_94}.
The quantum critical point was determined within this approximation.
Later, a Ginzburg-Landau theory \cite{sach_re_op_95}, which kept only terms relevant
or dangerously irrelevant under the quantum dynamical renormalization group,
was applied to study the critical exponents. 

Within this framework it turned out
that the quantum-dynamical Gaussian result for the shift exponent of the
critical temperature disagreed with the static approximation result due to the effect
of dangerously irrelevant dynamic couplings. The shift exponent $\phi$ can be defined
by $T_{c}(x)\sim(x_c-x)^{\phi}$, $x_c>x$, where $x$ stands for the parameter which
drives the quantum phase transition, and $T_c(x_c)$ vanishes by definition
at the quantum critical point (QCP).

Naturally, renormalization groups are not designed to determine critical points.
However, the positions of the QCPs are relevant, too. In finite dimensions the magnetic transition
may coincide with a localization transition driv\-en by the randomness of the
many-body interaction.
If these transitions do not coincide it is still important to know whether or not the
magnetic transition occurs within the metallic phase and, perhaps, to identify
parameters which influence the relative position of the QCPs. The answer may
in some detail be model-dependent, but a coincidence of both transitions would
almost certainly contain deeper reasons and could be expected to appear at least
in model classes.

In this work we adapt and apply the coherent potential approximation (CPA)
to a metallic spin glass problem. Originally this powerful non-perturbative
method was developed to describe non-interacting disordered electron systems
\cite{elli_krum_leath_74_CPA}, and in this context the CPA can be shown to
become exact in the limit of infinite spatial dimensions ($d\to\infty$)
\cite{vla_voll_92_cmf}. Later, the CPA formalism was generalized to deal with
interacting electron systems \cite{janis_89_CPA,janis_voll_92_coupling} with highly
non-trivial couplings of the Matsubara frequencies. Finally, the CPA method
can also be applied to dynamical disorder. Like in the present work, this situation
for instance results from the dynamical decoupling of interaction terms \cite{kake_91_MC}.

It is interesting to note that effective action terms that are non-diagonal in imaginary time not only occur in the presence of disorder. For certain translationally invariant models, the so-called extended DMFT approach \cite{smith_si_00_EDMFT}, which to some extend incorporates spatial correlations, gives rise to effective impurity problems equivalent to the one studied in this article.

The present article is organized as follows. The required technical tools and strategies
are developed in Sec. \ref{sec:derivation}. After the model definition we give a
brief description of the dynamical two-step decoupling procedure, which reduces the
problem to one of free fermions moving in a dynamical random magnetic field.
The model assumption of a fully connected magnetic interaction facilitates
a saddle point treatment. Extending previous work \cite{op_mg_93,op_bi_94}
we choose a {\it dynamical} self-consistent saddle point for the spin glass field.

The quantum-dynamical CPA method, shaped to apply to the present quantum spin
glass model, is discussed in Sec. \ref{sec:dyn_CPA}.
Within this framework, a set of  replica-symmet\-ric \seco equations for the
saddle point values $q$ (spin glass order parameter) and $\qt_m=\qt(\omega_m)$
(the local replica diagonal spin correlation) are derived.
As the centerpiece we obtain a matrix CPA-equation in Matsubara frequency space.

Our ideas for the approximate solution of these \seco equations are introduced in
Sec. \ref{sec:sol_strats}. As a systematic approximation scheme we propose to
restrict the dynamics of the effective random medium to a number of bosonic
Matsubara frequencies that can be numerically dealt with. Similar approximations
have been constructed earlier in the context of the Ising spin glass in the transverse field by means of discretization of the imaginary time axis \cite{usad_buet_90b,gold_lai_90}.

Section \ref{sec:static_approx} revisits the spin-static approximation where a
crucial simplification arises: the matrix structure of the equations disappears.
In particular, the CPA-equation can be solved independently for each Matsubara frequency. However, via the
saddle point values, the non-trivial coupling of the frequencies is preserved
also in the static set of self-consisten\-cy equations.  We present numerical solutions
for all temperatures including $T=0$. The spin-static $T=0$ critical point well agrees
with the results for an earlier model version \cite{op_bi_94}.

The main results follow in Sec. \ref{sec:qd_sol}. We evaluate the critical line
in the $T$--$t$ plane ($t$ represents the hopping strength) as a sequence of improved
dynamical approximations, which helps to derive the decay of the critical temperature
towards the quantum critical point. 
This dynamical approximation scheme is not designed to capture the zero-temperature limit. But, by increasing the number of
Fourier components of the effective random medium, one finds the $T_c$-deviation
from the spin-static approximation result almost accurately down to lower and
lower temperatures. 
It is seen that the  higher-frequency corrections to $T_c$
are small; their infinite number accumulates and leads to the non-analytical
behavior of the $T_c$-curve for $T_c\rightarrow0$.
From the characteristic decay of these corrections we deduce the location of the QCP.
The numerical results also fit with the quantum-dynamical shift exponent of the
$T_c$-curve and thus provide a reliable estimate of the QCP's position.


\section{Effective action and construction of the self-consistency method}
\label{sec:derivation}
\subsection{Model and spin glass decoupling procedure}
\label{sec:mo_def}
We consider the grand canonical Hamiltonian
\begin {equation}
\label{eqn:Hamiltonian} {\cal{K}}=\frac 12 \sum_{i\neq j}   J_{ij}
S^z_i S^z_j \wideop{-} \mu \sum_{i\sigma} a^\dagger_{i\sigma}
a_{i\sigma}\wideop{+}\ts\sum_{\langle ij\rangle\sigma}
a^\dagger_{i\sigma} a_{j\sigma}
\end{equation}
with the fermionic Ising spin operators given by
$S^z_i=a^\dagger_{i\uparrow} a_{i\uparrow}\wideop{-}
a^\dagger_{i\downarrow} a_{i\downarrow} $. The sum index $\langle
ij \rangle$ in the hopping term denotes summation over nearest-neighbor lattice sites. We assume quenched Gaussian disorder among the magnetic coupling constants $J_{ij}$ according to the distribution
\begin{equation}
\label{eqn:disorder_distribution}
P(J_{ij})=\frac{1}{\sqrt{2\pi}\Js}\expbb{-\frac{J_{ij}^2}{2\Js^2}}.
\end{equation}
In distinction to previous work
\cite{op_mg_93,op_bi_94,ros_op_98_mit} there is no disorder in the
kinetic part of \refeqn{eqn:Hamiltonian}. In order to facilitate
the solution of this model we assume a fully connected magnetic
interaction among the $N$ lattice sites. The hopping takes place
on an underlying simple cubic lattice in the limit of infinite
spatial dimensions $d$. To obtain physically meaningful results we
apply the usual scaling of the model parameters \cite{wolff_83}
\begin{equation}
\label{eqn:scaling}
\Js=J/\sqrt{N}\shb{0.5cm}{and}  \ts=t/\sqrt{d}.
\end{equation}

To some extent the derivation of the \seco equations of the
present model follows the detailed discussion given in
\cite{op_mg_93}. In the following we restrict ourselves to a brief
outline of the two-step decoupling procedure in the replica
formalism.

We start from the Grassmann field theoretic formulation of the
$n$-fold replicated partition function. The disorder average, i.e.
integration over the magnetic coupling constants with the Gaussian
weight \refeqn{eqn:disorder_distribution} generates four-spin
products in the effective action. Due to the assumed complete
connectivity of the magnetic interaction these four-spin products
can be reduced to quadratic terms by means of site-global, replica
and imaginary-time dependent real decoupling fields $\Qabfull$.
The disorder averaged replicated partition function at this stage
reads
\begin{equation}
\label{eqn:Z1}
\avdo{}{Z^n}=c^n   \pathintQ   \pathintPsi  \smsp \e^{-\Az-\At-\AJ}
\end{equation}
with the action terms
\begin{eqnarray}
\Az&=&\sum_{i\alpha\sigma}\intcond{\tau}\GFa{\ia\sigma}{\tau}
\round{\partial _\tau-\mu}\GF{i\alpha\sigma}{\tau}, \\
\At&=&\ts   \multsum{(ij)}{\alpha\sigma} \intcond{\tau}
\GFa{\ia\sigma}{\tau} \GF{j\alpha\sigma}{\tau},\\
\label{eqn:AJ} \AJ&=&\frac {J^2}{4}\intcond{\tau\tau'}
\multsum{\alpha,\beta}{i}\round{\round{\Qabfull}^2 -2\Qabfull
\smsp \GS{\ia}{\tau}\GS{i\beta}{\tau '}}.
\end{eqnarray}
Here $\GS{\ia}{\tau}=\sum_\sigma\GFa{\ia\sigma}{\tau}  \sigma
\GF{i\alpha\sigma}{\tau}$ is the time dependent Grassmann
representation of an Ising spin operator and the
$\tau$-integrations extend from $0$ to the inverse temperature
$1/T$.

The further evaluation of \refeqn{eqn:Z1} relies on the
elimination of the fields $\Qabfull$ by means of a saddle point
integration. The simplest but by no means trivial way to proceed
would be the assumption of a replica-symmetric and static (i.e.
$\ttp$-independent ) saddle point \cite{op_mg_93,op_bi_94}. Since
we want to explore however the role played by the quantum dynamics,
we have to keep the time dependence.

Let us symbolize the quantum statistical average and the
disorder average by
$\avth{}{\,}$ and $\avdo{}{\,}$, respectively.
Then, the saddle point matrix can be expressed in terms
of the corresponding averaged spin products:
\begin{eqnarray}
\label{eqn:sp_time_dependence_off_diag}
\left.\Qab\right|_{\mbox{s.p.}}&=&\avdo{}{\avth{\aneqb}{\GS{\ia}{\tau}\GS{i\beta}{\tau '}}}
= \avdo{}{\avth{}{\GS{\ia}{0}}\avth{}{\GS{i\beta}{0}}} \\
\label{eqn:sp_time_dependence_diag}
\left.\Qaa\right|_{\mbox{s.p.}}&=&\avdo{}{\avth{}{\GS{\ia}{\tau}\GS{\ia}{\tau '}}}
=\avdo{}{\avth{}{\GS{\ia}{\tau-\tau '}\GS{\ia}{0}}}.
\end{eqnarray}
Clearly, the inter-replica spin correlations are independent of
time because the fermions can not propagate between different
replications of the system.  All quantum-dynamical behavior of the
model originates from the diagonal elements of the saddle point
matrix. 

In this publication we choose a global replica-symmetric
saddle point (which is approximate only below $T_c$, but does
not affect the $T_c$-result of the second order phase transition)
with the appropriate time dependence according to
(\ref{eqn:sp_time_dependence_off_diag}, \ref{eqn:sp_time_dependence_diag}),
\begin{equation}
\label{eqn:sp}
\left.\Qab\right|_{\mbox{s.p.}}=q \shb{0.5cm}{and} \left.\Qaa\right|_{\mbox{s.p.}}=\qt_{|\tau-\tau'|}.
\end{equation}

We prefer to work in frequency space and perform Fourier
transformations of the Grassmann variables as well as of the
diagonal part of the saddle point matrix which take the form
\begin{eqnarray}
\label{eqn:FT}
\GF{}{\tau}&=&T \sumii{l}  \GF{}{l}  \smsp \e^{-i z_l \tau},\\
\qt_{|\tau-\tau'|}&=&\sum_{m=-\infty}^\infty  \qt_{m}  \smsp
\e^{-i\omega_m\dt},
\end{eqnarray}
where $z_l$ and $\omega_m $ denote fermionic and bosonic Matsubara
frequencies, respectively. The Fourier coefficients $\qt_m$ are
real quantities and obey the symmetry relation
$\qt_{m}=\qt_{-m}=\qt_{m}^*$.

In order to facilitate the Grassmann integration the effective
action must be further reduced to a quadratic form of the
Grassmann variables which requires another decoupling step.
Without going into detail we hence introduce Gaussian integrations
over site-local and replica-global fields $z_i$ and site-local and
replica-global fields $y_{\ia}$. The latter again split into
$\yzfull{\ia}$, which are also present in the spin-static theory, and
dynamical decoupling fields $\ymfull{\pm}{\ia}{m\geq 1}$, which comprise
the quantum dynamical character of the model. 

{\it Introduction of a space saving definition.} Throughout this article we will use the shorthand notation of the Gaussian integral operator
\begin{equation}
\label{eqn:GI}
\intGauss{x}  f(x)=\frac{1}{\sqrt{2\pi}}\int_{-\infty}^{\infty}
\rd x \smsp  \expbb{-\frac{x^2}{2}}  f(x)
\end{equation}
which renders many of the equations more compact.
\\

Employing this abbreviation we arrive at the completely decoupled partition function
\begin{equation}
\label{eqn:Z_decoupled}
\avdo{}{Z^n}=c^n\prod_i\intGauss{z_i}\prod_{\ia}
\intGauss{\yzfull{\ia}}\multprod{\ia}{m\geq 1}
\intGauss{\ymfull{+}{\ia}{m}}\intGauss{\ymfull{-}{\ia}{m}}
\pathintPsi   \smsp \e^{-\mathcal{A}_{\mbox{\tiny eff}}}
\end{equation}
with the effective action
\begin{eqnarray}
\label{eqn:Aeff}
\Aeff&=&\frac {nN J^2}{4T^2}\round{\sum_m \qt_m^2-q^2} +\At\nonumber\\
&&-T\sum_{\ia \sigma l}\round{iz_l+\mu+\sigma \Hzfull{\ia}}\GFaGF{\ia\sigma}{l}\\
&&-T\multsum{\ia \sigma l}{m\geq 1}\sigma\round{\Hmfull{\ia}{m}\GFa{\ia\sigma}{l+m}
\GF{\ia\sigma}{l}\wideop{+}{\Hmfull{\ia}{m}}^*\GFa{\ia\sigma}{l-m}\GF{\ia\sigma}{l}}.
\nonumber
\end{eqnarray}
In Eq. (\ref{eqn:Aeff}) we discarded irrelevant terms $\sim n^2$. The
occurring effective magnetic fields, which are complex, dynamical, and local
in site- and replica-indices, are given by
\begin{eqnarray}
\label{eqn:H0_eff}
\Hzfull{\ia}&=&J\round{\sqrt{q}\smsp z_i\wideop{+}\sqrt{\qt_0-q}
\smsp \yzfull{\ia}},\\
\label{eqn:Hm_eff}
\Hmfull{\ia}{m\geq1}&=&J\sqrt{\frac{\qt_m}2}\round{\ymfull{+}{\ia}{m}
+i\ymfull{-}{\ia}{m}}.
\end{eqnarray}


\subsection{The dynamical CPA approach}
\label{sec:dyn_CPA}
According to Eqs. (\ref{eqn:Z_decoupled}, \ref{eqn:Aeff}) the
problem has been reduced to an ensemble of non-interacting
fermions moving in a complex replica- and spin-dependent effective
random me\-dium. This situation immediately calls for a dynamical
version of the CPA \cite{kake_91_MC}. Following the prescription
of this method we replace the complex random medium $
\Hmfull{\ia}{m}$ by a yet unknown self energy $\Sigma_l$. The
chosen limit of infinite spatial dimensions simplifies the problem
to a single site problem and justifies the assumption of a
site-diagonal (or $k$-independent) self-energy
\cite{vla_voll_92_cmf,geor_etal_96_DMFT}.

The effective action \refeqn{eqn:Aeff} is not diagonal in the
energy indices thus allowing for virtual absorption and emission
of dynamical field quanta $\Hmfull{\ia}{m\geq1}$. However, the
full fermion Green's function of the original interacting problem
with any realization of the quenched disorder is certainly energy
conserving and so is the full disorder averaged Green's function.
Hence its off-diagonal elements in frequency space must vanish due
to the average over the effective random medium. There is also no
ferromagnetic tendency in the system which altogether justifies
our Ansatz of a spin-independent and frequency-diagonal self-energy.

In order to determine $\Sigma_l$ self-consistently we keep the
random medium at one single site, say $i=0$. The effective action
then reads
\begin{eqnarray}
\label{eqn:Acpa}
\Aeff^{CPA}&=&\frac {n N J^2}{T^2} \sum_m\round{\qt_m^2-q^2}\\
&&{}-T\multsum{ij}{\alpha\sigma
l'l}\GFa{\ia\sigma}{l}\round{\round{\Gmat{}^{-1}}_{ij}^{l'l}-\V{\alpha\sigma}{l'l}
\delta_{i0}\delta_{j0}   } \GF{j\alpha\sigma}{l}\nonumber
\end{eqnarray}
with the inverse of the full disorder averaged Green's function
\begin{equation}
\label{eqn:Ginv} \round{\Gmat{}^{-1}}_{ij}^{l'l}=\round{\round{i
z_l+\mu-\se{l}}\delta_{ij}  + \ts \smsp \delta_{\langle
ij\rangle}}\delta_{l'l}
\end{equation}
and the effective dynamical potential at the special lattice site $i=0$
\begin{equation}
\label{eqn:V}
\V{\alpha\sigma}{l'l}=\left\{
\begin{array}{cl}
\sigma \Hmfull{0\alpha}{m},& l'=l+m, m>0\\
\sigma \Hzfull{0\alpha}+\Sigma_l,\smsp &l'=l\\
\sigma \Hmfull{0\alpha}{m}^*,& l'=l-m, m>0.
\end{array}
\right.
\end{equation}

In the assumed case of infinite spatial dimensions near\-est-neighbor hopping of non-interacting particles is described by the
function (recall the scaling (\ref{eqn:scaling}))
\begin{equation}
\label{eqn:T} \Tfree(x,t)=-\sign{\round{\Im \smsp x }}\frac
{\sqrt{\pi}}{2t} \expbb{-\frac{x
^2}{4t^2}}\round{i+\erfi{\frac{x}{2t}}}
\end{equation}
which reveals for $x=\varepsilon+i0^+$ the well-known Gaussian
density of states \cite{wolff_83,muell_89_corr_fer}. Hence the
full Green's function $\Gmat{}$ defined in \refeqn{eqn:Ginv} is
readily expressed by
\begin{equation}
\label{eqn:GbyT}
\round{\Gmat{}}_{ij}=\Tmat \delta_{ij}\wideop{+}\order{\frac{1}{\sqrt{d}}}
\end{equation}
where $\Tmat$ is site independent and diagonal in the Matsubara indices:
\begin{equation}
\label{eqn:Tmatdef}
\round{\Tmat}_{l'l}=\Tfree\round{i z_l+\mu-\Sigma_l,t}\delta_{l'l}.
\end{equation}

For the formulation of the self-consistency equations it is useful
to define as an auxiliary quantity the site-local propagator at
site $i=0$ in the presence of the effective potential,
\begin{equation}
\label{eqn:Glocdef} \round{\Glocmat{\alpha\sigma}}_{l'l}=T
\avAloc{}{\GFa{0\alpha\sigma}{l}\GF{0\alpha\sigma}{l'}}
=\round{\squared{\Tmat^{-1}\wideop{+}
\Vmat{\alpha\sigma}}^{-1}}_{l'l}.
\end{equation}
In essence, within the CPA method the frequency-depend\-ent self-energy $\Sigma_l$ is determined by the demand that the average of
the local propagator matrix $\Glocmat{\alpha\sigma}$ at the
special site $i=0$ with respect to the dynamical potential
\refeqn{eqn:V} coincides with the local part of the full  disorder
averaged homogeneous Green's function, $\Tmat$. Note that equating
both quantities requires the replica limit $\replim$ to be taken.

Omitting the algebraic details we obtain the conditional matrix
equation (to condense the notation all superfluous site-, replica-,
or spin-indices are dropped from now on)
\begin{equation}
\label{eqn:CPAequation}
\Tmat=\avzint{\avyint{\Glocmat{\sigma}}}
\end{equation}
where the symbol $\avyint{\,}$ is a shorthand notation for the
average with respect to all replica-local decoupling fields
$\vec{y}=\{\ym{}{0},\ym{+}{m},\ym{-}{m}  \}$,
\begin{equation}
\label{eqn:avyintdef} \avyint{f(z,\vec{y})}=\frac
{\intGauss{\ym{}{0}}\prod_{m\geq
1}\intGauss{\ym{+}{m}}\intGauss{\ym{-}{m}} \weightP(z, \mathbf{y})
f(z,\vec{y})} {\intGauss{\ym{}{0}}\prod_{m\geq
1}\intGauss{\ym{+}{m}}\intGauss{\ym{-}{m}}\weightP(z,
\mathbf{y})}.
\end{equation}
\begin{figure}
\begin{center}
\epsfig{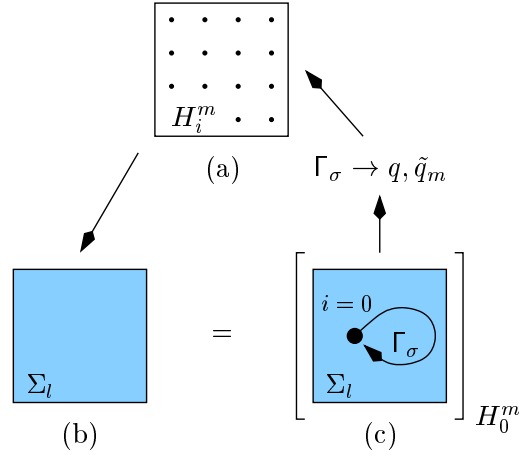}
\caption{Structure of the \seco problem of the itinerant spin
glass model (inspired by \cite{janis_voll_92_coupling}). (a)
indicates the effective dynamical random medium $\Hmfull{i}{m}$,
which depends on the spin glass order parameter $q$ and the
replica-diagonal saddle point $\qt_m$ (Eqs.
(\ref{eqn:H0_eff}, \ref{eqn:Hm_eff})). (b) and (c) illustrate the
determination of the homogeneous self-energy $\Sigma_l$ by the
CPA-equation (\ref{eqn:CPAequation}). From the local propagator
$\Glocmat{\sigma}$ at the special lattice site $i=0$ the physical
quantities $q$ and $\qt_m$ are constructed, which in turn generate
the original random medium in (a).} \label{fig:sc_structure}
\end{center}
\end{figure}
The weight function $\weightP(z, \mathbf{y})$ that appears in
\refeqn{eqn:avyintdef} results from the integration of the
Grassmann fields and is given by ($\tr{l\sigma}$ comprises spin
and frequency summation)
\begin{equation}
\label{eqn:weightP_def} 
\weightP(z,\mathbf{y})=\expbb{\tr{l\sigma}\ln\squared{1+\Vmat{\sigma}\Tmat{}}}.
\end{equation}
Note that, as discussed above, the off-diagonal elements of
$\Glocmat{\sigma}$ as well as its spin dependence vanish exactly
by integration.

Expressions for the saddle point values \refeqn{eqn:sp} can be
obtained by construction of the spin products given by Eqs.
(\ref{eqn:sp_time_dependence_off_diag},\ref{eqn:sp_time_dependence_diag})
in terms of the Grassmann fields at the special lattice site $i=0$
and application of Wick's theorem. After performing the Fourier
transformation \refeqn{eqn:FT} and taking the replica limit
$\replim$ the set of \seco equations is thus completed by
\begin{eqnarray}
\label{eqn:qdef}
q&=&T^2\avzint{\round{\avyint{\tr{l\sigma}\sigma\Glocmat{\sigma}}}^2}\\
\label{eqn:qtmdef}
\qt_m&=&T^2 \avzint{\multsum{ll'}{\ssp}\ssp \\
&&{}\times\avyint{\Gloc{\sigma}{l+m,l}\Gloc{\sigma'}{l'-m,l'}
-\Gloc{\sigma}{l+m,l'+m}\Gloc{\sigma'}{l',l}\delta_{\ssp}}}.\nonumber
\end{eqnarray}
In \refeqn{eqn:qtmdef} the first term involves the product of
sums over the $m^\mathrm{th}$ super- and sub-diagonals and the
second the trace of the matrix product of two factors
$\Glocmat{\sigma}$ shifted against each other about $m$ elements
along the diagonal.

Note that in the limiting case of a vanishing hopping strength
($t\to 0$) Eqs. (\ref{eqn:qdef}, \ref{eqn:qtmdef}) correctly
recover the results of the non-itinerant model \cite{op_mg_93} and
with the additional choice of $\mu=-i\pi T/2$ \cite{popo_fedo_88}
the equations further reduce to the SK-solution of the classical
model \cite{kirk_sher_78_sksol}.


\subsection{General solution strategies}
\label{sec:sol_strats}
Any attempt to solve the set of \seco equations
(\ref{eqn:CPAequation}, \ref{eqn:qdef}, \ref{eqn:qtmdef}) faces
the fundamental problem of the infinitely many quantities $\qt_m$
each of which effectuates corresponding Gaussian integrations via
Eqs. (\ref{eqn:H0_eff}, \ref{eqn:Hm_eff}, \ref{eqn:weightP_def}). In
order to render the problem feasible we propose to keep only a few
Fourier components $\qt_m$  with $m=\{0,\ldots , M \}$ and take
the higher-frequency components to be zero, i.e. $\qt_{m>M}\equiv
0$. In turn this means that the Gaussian integrations associated
with the components  $\qt_{m>M}$ become trivial. Below we will
refer to this approximation scheme as the ``dynamical
approximation of order $M$''. 

Within this approximation scheme the quantum dynamics is treated on energy scales ranging from $\omega_0=0$ to $\omega_M=2\pi T M$. To estimate the quality of this approximation one has to compare the energy scales that are neglected to the hopping strength $t$ as the model parameter that generates the quantum dynamics. Thus we are led to 
\begin{equation}
t\ll\omega_{M+1}\equiv 2\pi T (M+1)
\end{equation}
as a simple criterion of validity of the $M^\mathrm{th}$ order dynamical approximation.
Hence, although it is neither a high temperature expansion nor an expansion in small $t$, the method works well especially for small $t/T$. In those regions in parameter space the approximation already at manageable low orders $M$ excellently captures the effects of the quantum dynamics

Another difficulty arises from the infinite extension of the
matrices $\Tmat$, $\Vmat{}$ (which becomes a band matrix with $M$
sub- and super-diagonals (\ref{eqn:V}) in the dynamical
approximation of order $M$)  and $\Glocmat{}$ in frequency space.
Naturally, a numerical analysis requires the restriction to finite matrices of size $2(\lc+1)\times
2(\lc+1)$, i.e. the matrices are constructed in the limited frequency range $z_{-l_c-1}$ to $z_{l_c}$. However, there are also important contributions from higher frequencies that can not be neglected for accurate solutions.  

We overcome this problem by systematic asymptotic expansions of
the \seco equations in terms of $1/z_l$ up to some feasible order $\order{(1/z_l)^K}$. Here we exploit the high-frequency asymptotics of
the self-energy
\begin{equation}
\label{eqn:Sigma_asymptotics} \Sigma_l
\wideop{\lrarrowsubtext{|l|\!\!\to\!\!\infty}
}\sum_{k=1}^{K}
a_k\round{iz_l}^{-(2k-1)}\wideop{+}\order{z_l^{-(2K+1)}}
\end{equation}
where the expansion coefficients $a_k$ are easy to calculate
averages of polynomials of the effective fields (\ref{eqn:H0_eff},
\ref{eqn:Hm_eff}). The sums over the Matsubara frequencies which
occur in Eqs. (\ref{eqn:weightP_def}, \ref{eqn:qdef}, \ref{eqn:qtmdef})
can always be split up into a low-frequency main part and a high-frequency part
which are separated by the cut-off index $\lc$. While the matrix-structured main
part has to be treated numerically the high-frequency
contributions can be formulated in terms of asymptotic series expansions of
docile structure that permits analytical summation.

The approximation of the high-frequency contributions by asymptotic series expansions introduces some error.
The cut-off index $\lc$ has to be chosen such that this error undershoots some given threshold of insignificance. 
In practical calculations we used  different methods to determine $l_c$. 
A simple way is to make trial variations of the matrix size at each iteration cycle and to adjust (increase or decrease) $l_c$ according to the corresponding variations of all relevant intermediate quantities. 
A more direct method is to evaluate the contributions of the first neglected asymptotic order, i.e. $\order{(1/z_l)^{K+1}}$, as a function of $l_c$ and to apply some suitable smallness criterion. The latter method turned out to be un-practical for $M>0$ because of the complexity of the occurring analytical expressions for the asymptotic Matsubara sums. 

The final criterion for $l_c$ is always that the physical quantities are independent of this auxiliary parameter at some desired level of precision. The proper choice of
$\lc$ and thereupon the computational expenses for solving the
\seco equations strongly depend on the temperature as well as on
the order $K$ up to which the asymptotic series expansions of the
equations can be driven.
\begin{figure}
\begin{center}
\epsfig{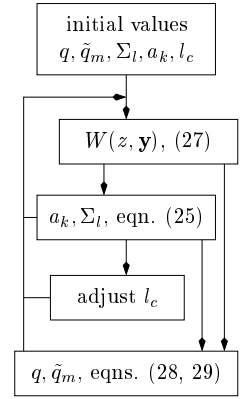}
\caption{Basic iterative scheme \cite{implementation} for the
solution of the coupled \seco equations (explanation in the
text).} \label{fig:iteration_scheme}
\end{center}
\end{figure}

All solutions of the \seco equations that are presented in this article
have been obtained by means of the principal iterative algorithm
sketched in Fig. \ref{fig:iteration_scheme}. This procedure proved
to be insensitive to the initial values and showed quite
satisfying convergence properties in all regions of the parameter
space explored so far.

For the sake of simplicity from now on we restrict ourselves to
the case of a vanishing chemical potential, $\mu=0$, which
corresponds to half fermion filling due to particle-hole-symmetry
of the Hamiltonian (\ref{eqn:Hamiltonian}). Without loss of generality we always set $J\equiv 1$.

\section{Spin-static approximation}
\label{sec:static_approx}
This section is devoted to a discussion of the dynamical \seco equations in the static approximation as the first and simplest of a sequence of the dynamical approximations proposed in the last section. This static approximation consists in neglecting the time dependence of the saddle point (\ref{eqn:sp}) or equivalently in taking all Fourier components $\qt_m$ with $m>0$ to be zero, i.e. $M=0$. This restriction to the static component $\qt_0$ implicates tremendous simplifications of the \seco equations. 

Because the dynamical effective fields \refeqn{eqn:Hm_eff} vanish the decoupling fields $\ym{\pm}{m}$ can be integrated out trivially. One is left with only two Gaussian integrations over the static fields $\ym{}{0}$ and $z$. Also, within this approximation the occurring matrices become diagonal and thus the matrix structure of the \seco equations disappears. Thus,  the dynamical CPA-equation \refeqn{eqn:CPAequation} decouples into a set of scalar equations for each Matsubara frequency that can be solved one at a time. Furthermore, the matrix inversion \refeqn{eqn:Ginv} turns into simple scalar inversion and the evaluation of the weight function \refeqn{eqn:weightP_def} reduces to an easily manageable numerical Matsubara product,
\begin{eqnarray}
\label{eqn:static_weight_P}
\weightP_{\mathrm{static }}(z,\ym{}{0})&=&\frac{1}{2} \round{\cosh\round{ \Hz{}/T}+1}   \\
&&\hspace{1cm}\times\prod_{l=0}^{\infty} \round{\frac{\absolute{u_l}^2+\Hz{}^2}{z_l^2+\Hz{}^2}}^2,\nonumber
\end{eqnarray}
where $\Hz{}=\sqrt{q} \smsp z+\sqrt{\qt_0-q}\smsp\ym{}{0}$. The first term in \refeqn{eqn:static_weight_P} is the suitable regularized frequency and spin  product of $i z_l+\sigma \Hz{}$ and constitutes the weight function of the non-itinerant model \cite{op_mg_93}. In the present itinerant model the effect of the hopping becomes noticeable in the deviation of $u_l$ from $iz_l$, where the first is defined by
\begin{equation}
\label{eqn:u_def}
u_l=\frac{1}{\Tfree(i z_l-\Sigma_l,t)}+\Sigma_l.
\end{equation} 
In terms of the functional
\begin{equation}
\label{eqn:A_finite_T}
A(x)=4x\sum_{l=0}^\infty \frac{1}{\absolute{u_l}^2+x^2}
\end{equation}
the expressions for the saddle point values $q$ and $\qt_0$ given in (\ref{eqn:qdef}, \ref{eqn:qtmdef}) simplify to
\begin{eqnarray}
\label{eqn:q_static}
q&=&   T^2 \avzint{\round{\avystaticint{A(\Hz{})}}}^2 ,\\
\label{eqn:qt_static}
\qt_0&=&  T^2 \avzint{\avystaticint{A(\Hz{})^2+A'(\Hz{})} }    .
\end{eqnarray}
\begin{figure}
\begin{center}
\epsfig{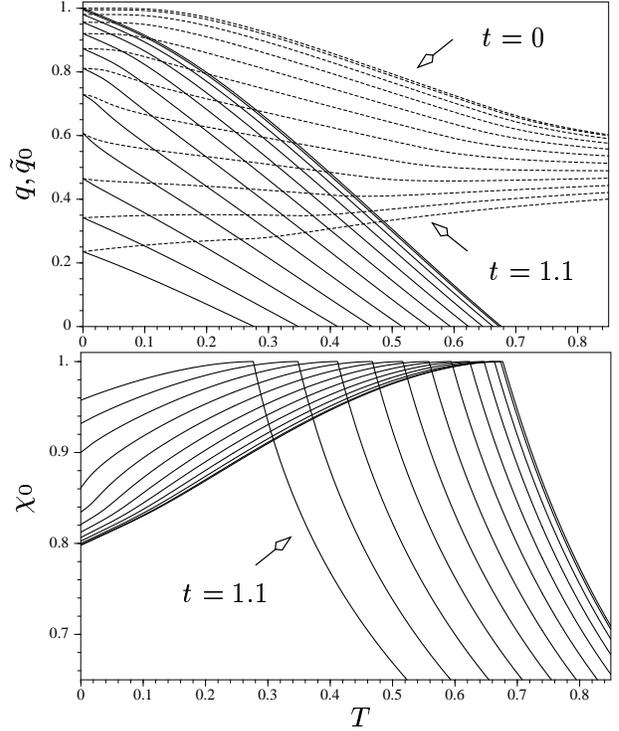}
\caption{Finite temperature results in the spin-static approximation for hopping strengths $t=0\ldots 1.1$ in steps of $0.1$. The upper plot shows the spin glass order parameter $q$ (full lines) and the zero frequency component of the replica diagonal saddle point value $\qt_0$ (dashed lines). The $\qt_0$-curves approach $1/2$ as $T\to\infty$. Below is plotted the corresponding local static susceptibility $\chi_0=(\qt_0-q)/T$ which remains finite as $T\to 0$. At the spin glass-paramagnet transition $\chi_0$ always reaches unity.}
\label{fig:qqt_static}     
\end{center}
\end{figure}

The numerical solutions of the static set of self-consist\-ency equations are presented in Fig. \ref{fig:qqt_static}. As $T\to\infty$ all available many particle states become equally populated and $\qt_0\to 1/2$ since two of the four local states are magnetic. In the non-itinerant limit $t=0$ the paramagnet to spin glass transition occurs at $T_c=1/\round{1+\expbb{-1/(2T_c)}} \simeq 0.6767$. The hopping hampers the local freezing of the spins and lowers the critical temperature. 


\subsection{The limit of zero temperature}
\label{sec:zero_temp}
\begin{figure}
\begin{center}
\epsfig{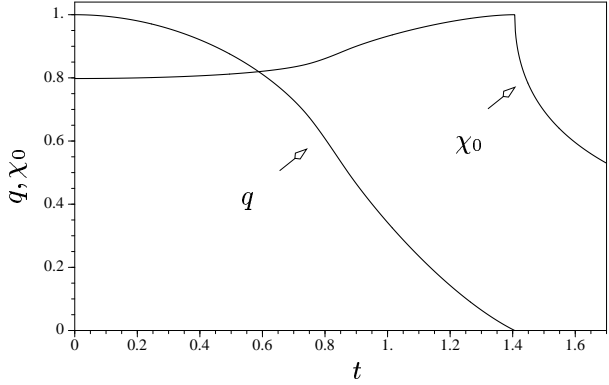}
\caption{Zero-temperature results for the spin glass order parameter $q=\qt_0$ and the local static susceptibility $\chi_0$. Increasing hopping strength depresses the spin glass order and drives a zero-temperature phase transition at $t_{cs}\simeq1.406$}
\label{fig:T=0_results}
\end{center}
\end{figure}

The numerical solutions of the static equations feature the low temperature behavior $(\qt_0-q)\sim T$. In order to perform the zero-temperature limit it is advisable to eliminate $\qt_0$ and to formulate the equations in terms of $q$ and the static part of the local susceptibility $\chi_0=(\qt_0-q)/T$ which remains finite as $T\to 0$ (Fig. \ref{fig:T=0_results}). We replace the discrete frequencies $z_l$ by the continuous variable $\zeta$ and recall the zero-temperature limit of the Matsubara summations,
\begin{equation}
\label{eqn:T_0_Matsubara_sum}
T \sum_{l=0}^\infty  f(z_l)\wideop{\lrarrowsubtext{T\!\!\to\!\!0}} \frac{1}{2\pi}  \int_0^\infty \rd \zeta \smsp f(\zeta).
\end{equation}
By means of the rescaling of the integration variable \linebreak $\sqrt{T}\smsp\ym{}{0} \to \ym{}{0}$ the weight function \refeqn{eqn:static_weight_P} together with the Gaussian factor assume the form $\expbb{- g(z,\ym{}{0})/T}$. The exponent function $ g(z,\ym{}{0})$ can be shown to remain finite as $T\to 0$. Thus, the $\ym{}{0}$-integration reduces to a simple saddle point integration where the $z$-dependent saddle point has to be determined numerically. 
Finally, we derive the following set of \seco  equations at zero temperature:
\begin{equation}
\label{eqn:T_0_CPA}
\Tfree\round{i\zeta-\Sigma(\zeta),t}=\frac 12 \intGauss{z}\sum_{\sigma=\pm 1}\frac 1{u(\zeta)+\sigma \eta(z)},
\end{equation}
\begin{equation}
\label{eqn:T_0_q}
q=\qt_0=\intGauss{z} \AzeroT^2(\eta(z))
\end{equation}
\begin{equation}
\label{eqn:T_0_chi}
\chi_0=\sqrt{\frac{2}{\pi q}}+\intGauss{z}\frac{\AzeroT'(\eta(z))}{1-\chi \AzeroT'(\eta(z))}
\end{equation}
\begin{equation}
\label{eqn:T_0_saddle_point_eqn}
\eta(z)=\chi \AzeroT(\eta(z))+\sqrt{q} |z|,
\end{equation}
with the functional
\begin{equation}
\label{eqn:functional_A}
\AzeroT(x)=\frac{2x}{\pi} \int_0^\infty\rd \zeta  \smsp  \frac{1}{\absolute{u(\zeta)}^2+x^2}. 
\end{equation}
In Eqs. (\ref{eqn:T_0_CPA}--\ref{eqn:functional_A}) $u(\zeta)$ and $\Sigma(\zeta)$ are continuous versions of the quantities $u_l$ and $\Sigma_l$ defined in Sec. \ref{sec:dyn_CPA} and (\ref{eqn:u_def}).
In terms of the constant
\begin{equation}
\label{eqn:constant_c}
\gamma=\int_0^\infty \rd \zeta \smsp \expbb{2 \zeta^2}\Gamma\round{\frac 12,\zeta^2}^2
\end{equation}
the static quantum critical point is located at $t_{cs}=2\gamma/\pi\backsimeq1.406$. Up to a slight deviation due to different model definitions this critical value quantitatively agrees with the results obtained in \cite{op_bi_94} for the case of a semi-elliptic free energy band.

In the zero-temperature disordered phase, i.e. for $t>t_{cs}$, the static part of the local susceptibility is given by $\chi_0(t)|_{T=0}=(\pi t-\sqrt{\pi^2t^2-4\gamma^2})/(2\gamma)$. Its deviation from the corresponding quantity in the non-interacting limit,  $\chi_0(t)|_{T=0, J=0}=\gamma/(\pi t)$, signalizes the vicinity of the spin glass phase.



\section{Quantum dynamical solutions}
\label{sec:qd_sol}
While exact in the non-itinerant limit, the spin-static approximation discussed in Sec. \refsec{sec:static_approx} turns out to yield a very good description of our model for weak and moderate hopping. For stronger hopping, however, this static approximation becomes increasingly inaccurate and particularly fails close to the $T=0$ quantum phase transition. In this section we present improved solutions of the \seco equations within the dynamical approximation of up to third order as introduced in Sec. \ref{sec:sol_strats}. 

For the time being we restrict ourselves to the determination of the phase diagram of the model. Since at criticality there is no issue of replica symmetry breaking the choice of the replica-symmetric saddle point \refeqn{eqn:sp} is justified in these calculations. We determine the critical curve by virtue of the exact relation
\begin{equation}
\label{eqn:Tc_det}
T_c=\qt_0\round{T_c}
\end{equation}
which was first derived in \cite{op_mg_93} in the static approximation and can be shown to hold within the present dynamical treatment, too (see \ref{sec:app_eqn}). Equation \refeqn{eqn:Tc_det} implies that the static part of the local susceptibility, $\chi_0$ reaches unity at the phase transition. In solving the conditional equation \refeqn{eqn:Tc_det} it is sufficient to fix the spin glass order parameter to $q=0$ thus rendering the $z$-integrations in Eqs. (\ref{eqn:CPAequation}, \ref{eqn:qtmdef}) trivial.
\begin{figure}
\begin{center}
\epsfig{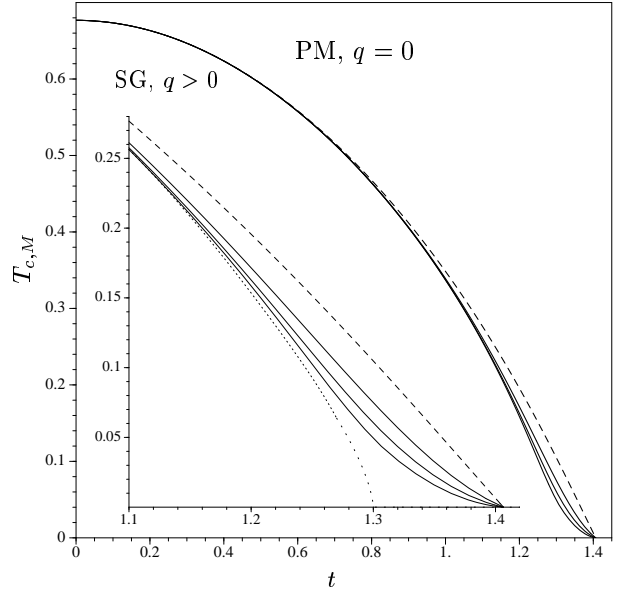}
\caption{The critical line of the spin glass (SG) to paramagnet (PM) phase transition in the static approximation (dashed line) and in the dynamical approximations (full lines) of first ($T_{c,1}$, uppermost) to third order as discussed in Sec. \ref{sec:sol_strats}. The dotted line indicates the expected fully dynamical phase boundary; the light-dotted part shows the pure critical behavior (\ref{eqn:critical_exponent}).}
\label{fig:phase_diagram}
\end{center}
\end{figure}

\begin{figure}
\begin{center}
\epsfig{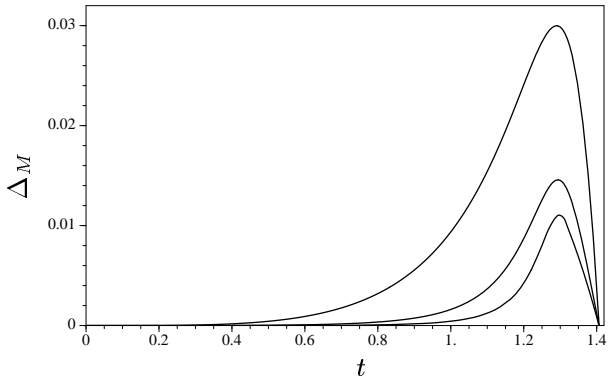}
\caption{Differences of the critical lines in two successive orders of the dynamical approximation, $\Delta_M=T_{c,M-1}-T_{c,M}$.}
\label{fig:diff_plot}
\end{center}
\end{figure}

Our solutions for the critical line $T_c(t)$ as a sequence of the first three orders of the dynamical approximation ($M=\{1,2,3\}$) are shown in Fig. \ref{fig:phase_diagram}. With increasing hopping strength and decreasing temperature the growing influence of the discrete dynamic saddle point components $\qt_{m>0}$  is getting more and more apparent. It can be seen clearly from Fig. \ref{fig:diff_plot} that with increasing order of the dynamical approximation two successive solutions start to separate at larger $t$. We observe a rapid convergence of this sequence of solutions except for the region where the quantum phase transition is expected. As $T_c\to 0$ all curves collapse into the static critical point at $t_{cs}$. 

In the disordered phase the $\qt_m$ and consequently the effective potential matrix \refeqn{eqn:V} vanish linearly with temperature, i.e. the dynamical susceptibility $\chi_m=\qt_m/T$ has a finite zero-temperature limit. Hence the self-consistent inclusion of any finite number of the $\qt_m$ can affect neither the position of the QCP nor the critical exponents. In order to capture the quantum dynamical character of the problem it is necessary to take into account the $\qt_m$ over a finite range of Matsubara frequencies $\omega_m$ around $\omega_0=0$. Thus, the linear temperature decrease of the $\qt_m$ is compensated by the increasing number of Fourier components within a fixed frequency range and the location of the critical point is shifted towards smaller hopping strength compared to the static approximation (Sec. \ref{sec:app_expansion}). 

Close to zero temperature the critical line behaves like
\begin{equation}
\label{eqn:critical_exponent}
T_c\sim(t_c-t)^\phi,\hspace{0.3cm}t<t_c
\end{equation}
with the shift exponent changing from $\phi=1$ in the static approximation to $\phi=2/3$ \cite{sach_re_op_95} due to the quantum dynamics (Sec. \ref{sec:app_expansion}). Figure \ref{fig:phase_diagram} gives an impression of how this non-analytical behavior emerges from the sequence of the (analytical) approximate solutions. 
The differences between two successive approximations, 
\begin{equation}
\label{eqn:DeltaM}
\Delta_M=T_{c,M-1}-T_{c,M}, 
\end{equation}
exhibit pronounced maxima (Fig. \ref{fig:diff_plot}). While the positions of these maxima vary only very little they become lower in height but sharper with increasing $M$.

The critical lines $T_{c,M}$ are monotonically decreasing functions of $t$. Hence the distance between $T_{c,M}$ and the fully dynamical true critical line, $\Delta^\infty_M=T_{c,M}-T_{c,\infty}$, possesses a non-analytical maximum exactly at $t_c$ for any $M$. Since
\begin{equation}
\label{eqn:sumrule}
\Delta^\infty_M=\sum_{M'=M+1}^\infty \Delta_{M'},
\end{equation}
this non-analyticity must coincide with the position of the maxima of the $\Delta_M$ as $M\to\infty$. 
This simply means that the sequence of the critical lines $T_{c,M}$ converges slowest in the very proximity of the QCP.
Based on this scenario we estimate the location of the QCP: we plot the maxima positions vs. their heights and extrapolate to zero height. This simple procedure yields the final result 
\begin{equation}
t_c\simeq 1.30 .
\end{equation}



\section{Summary and outlook}
\label{sec:conclusion}
We considered a fermionic spin glass model including a nearest-neighbor hopping term. By means of standard decoupling techniques the problem has been reduced to that of a dynamical random field system. A set of \seco equations for the spin glass order parameter $q$ and the Fourier components of the replica-diagonal saddle point $\qt_m$ has been derived by virtue of a dynamical CPA method.
In order to facilitate numerical solutions we kept only a manageable number of low frequency components $\qt_{m\leq M}$ in the equations and abstain from the self-consistent evaluation of the $\qt_{m>M}$. We referred to this scheme as the dynamical approximation of order $M$. 

Within the static approximation ($\qt_{m>0}=0$) we presented solutions both at finite and zero temperature. The second order $T=0$ phase transition was found at $t_{cs}=2\gamma/\pi\simeq1.406$, Eq. (\ref{eqn:constant_c}). In order to determine the phase diagram of the model we calculated the SG-PM phase boundary in the dynamical approximation in up to third order (Fig. \ref{fig:phase_diagram}). These data allowed to estimate the location of the fully dynamical critical point at $t_c\simeq 1.30$.
In order to confirm this result it would be desirable to find solutions in higher orders of the dynamical approximation, i.e. $M>3$. 

In this article we concentrated on the spin sector of our model and left out the properties in the charge sector such as the fermionic density of states. In the future it will be of high interest to investigate the effect of the hopping on the band structure of the system, particularly on the spin glass gap at zero temperature \cite{op_ros_98_gap}. In this context an extension of the solutions to non-zero chemical potential $\mu$ \cite{ros_op_96} is also desirable, too.

There are also important questions concerning the interplay between quantum dynamics and replica symmetry breaking \cite{op_ros_98_rsb}, both issues being most significant  at $T=0$.
\\

This work was supported by the Deutsche Forschungs\-gemeinschaft under research project Op28/5--2 and by the SFB410. One of us (M. B.) also wishes to acknowledge  the scholarship granted by the University of W\"urzburg.


\begin{appendix}
\section*{Appendix}
\setcounter{section}{1}
\subsection{Derivation of Eq. (\ref{eqn:Tc_det})}
\label{sec:app_eqn}
In order to locate the spin glass phase transition we expand Eq. (\ref{eqn:qdef}) in terms of $q$ (note that $q$ enters the equations only in the combination $\sqrt{q}z$ by Eq. (\ref{eqn:H0_eff})). With the abbreviation $A(z,\vec{y})=\tr{l\sigma}\sigma\Glocmat{\sigma}(z,\vec{y})$ we have
\begin{eqnarray}
\label{eqn:qdef_expansion}
q&=&T^2\avzint{\left(\avyintzo{A} \wideop{+}\left.\partial_{\sqrt{q}z}\avyint{A}\right|_{q=0}\sqrt{q}z\right)^2}\nonumber\\
&&\ \wideop{+}\order{q^2}.
\end{eqnarray}
Recall the definition of the average $\avyint{\,}$, Eq. (\ref{eqn:avyintdef}).
The symmetry relations $A(0,\vec{y})=-A(0,-\vec{y})$ and $\weightP(0,\mathbf{y})= \weightP(0,-\mathbf{y})$ readily follow from Eqs. (\ref{eqn:H0_eff}, \ref{eqn:Hm_eff}, \ref{eqn:V}, \ref{eqn:Glocdef}). Hence, the first term in Eq. (\ref{eqn:qdef_expansion}) vanishes by the $\vec{y}$-integration.

For the second term in Eq. (\ref{eqn:qdef_expansion}) we need to evaluate
\begin{equation}
\label{eqn:dWA_chainrule}
\partial_{\sqrt{q}z} WA=\round{\partial_{\sqrt{q}z} W}A\wideop{+}W\partial_{\sqrt{q}z} A.
\end{equation}
We expand the terms $\ln\squared{1+\Vmat{\sigma}\Tmat{}}$ and $\squared{\Tmat^{-1}\wideop{+}\Vmat{\sigma}}^{-1}$ that occur in the expressions for $W$ and $A$, respectively (Eqs. (\ref{eqn:weightP_def}, \ref{eqn:Glocdef})), in powers of the matrix $\Vmat{\sigma}$ (\ref{eqn:V}). After taking the derivative these series can be re-summed easily yielding
\begin{eqnarray}
\label{eqn:dW}
\partial_{\sqrt{q}z} W&=&W A,\\
\label{eqn:dA}
\partial_{\sqrt{q}z} A&=&-\tr{l\sigma}\Glocmat{\sigma}^2.
\end{eqnarray}
Altogether, very close to the phase transition where, $q\simeq 0$, Eq. (\ref{eqn:qdef_expansion}) reads
\begin{equation}
\label{eqn:Tc_det_aux}
1=T^2 \avzint {z^2 \,\round{ \avyintzo{\round{\tr{l\sigma}\sigma\Glocmat{\sigma}}^2\wideop{-}\tr{l\sigma}\Glocmat{\sigma}^2}}^2}.
\end{equation}
The remaining Gaussian $z$-integration evaluates to $1$.  By comparing to Eq. (\ref{eqn:qtmdef}) at $T=T_c$, the right hand side of Eq. (\ref{eqn:Tc_det_aux}) can be identified with $\qt_0^2/T_c^2$ which finally proves relation (\ref{eqn:Tc_det}). 

\subsection{Expansion in small $\qt_m$}
\label{sec:app_expansion}
In order to extract the behavior of the critical line $T_c(t)$ at the quantum phase transition we expand the \seco equations in powers of the effective potential matrix $\V{\sigma}{}$ \refeqn{eqn:V} in the disordered phase, i.e. for $q=0$. The internal summations due to the occurring matrix multiplications compensate the linear temperature decrease of the Fourier components of the dynamic saddle point, $\qt_m$. An expansion up to second order  yields the following \seco equation for the local dynamical susceptibility $\chi_m=\qt_{|m|}/T$:
\begin{eqnarray}
\label{enq:chi_dspe}
\chi_m&=&C_m\round{1+\chi^2_m}\\
&&{}-\round{1-C_m\chi_m}T\sum_{n=-\infty}^\infty \frac{F_{m,n}\chi_{n}}{1-C_{n}\chi_{n}.}\nonumber
\end{eqnarray}
Using the abbreviation $X_l=1/\round{1/T_0(i z_l-\Sigma_l,t)+\Sigma_l}$ (see \refeqn{eqn:T}) the coefficients that occur in Eq. (\ref{enq:chi_dspe}) are defined by 
\begin{eqnarray}
\label{eqn:C}
C_m&=&-2T\sum_{l=-\infty}^\infty X_lX_{l+ m},\\
\label{eqn:F}
F_{m,n}&=&T\sum_{l=-\infty}^\infty X_lX_{l+ m}X_{l+n}\round{2X_l+X_{l+m+n}},
\end{eqnarray}
and the self-energy is given explicitly by
\begin{eqnarray}
\label{enq:Sigma_expansion}
\Sigma_l&=&\frac 12 X_l^2 T \sum_{m=-\infty}^\infty\frac{\chi_m X_{l+m}}{1-C_m\chi_m}.\\
&&{}\nonumber
\end{eqnarray}

By  numerical evaluation of Eq. (\ref{enq:chi_dspe}) in the zero-temperature limit we determine a critical hopping strength $t_c\simeq 1.372$ well between the static result and the expected fully dynamical value given in Sec. \ref{sec:qd_sol}.

Expansion of Eqs. (\ref{enq:chi_dspe}--\ref{eqn:F}) in the hopping strength around $t_c\equiv t_c(T=0)$ and in small frequencies $\omega_m$ finally yields the condition for the critical line at small temperatures
\begin{eqnarray}
\label{eqn:exp_det_eqn}
t_c-t_c(T)&\sim&T\sum_{\omega_m=0}^{\omega_\Lambda}\sqrt{\omega_m}-\left.T\sum_{\omega_m=0}^{\omega_\Lambda}\sqrt{\omega_m}\right|_{T\to 0}\nonumber\\
&\sim&{}T^{3/2}.
\end{eqnarray}


\end{appendix}


\begin{thebibliography}{10}
\bibitem{implementation}
{The basic algorithm was implemented with the Wolfram Mathematica language
  while the computationally expensive matrix operations where handled by
  external C routines making use of the MathLink interface.}
\end{thebibliography}
\end{document}